\documentclass[floatfix,twocolumn,nofootinbib,superscriptaddress,a4paper,preprintnumbers]{revtex4-1}

\usepackage{amsmath}
\usepackage{amssymb}
\usepackage{graphicx}
\usepackage[T1]{fontenc}
\usepackage{slashed}
\usepackage[utf8]{inputenc}
\usepackage{xspace}
\usepackage{color}
\usepackage{ulem}
\usepackage{array}

\newcommand{\AddrYantai}{Department of Physics, Yantai University, Yantai 264005, P. R. China}
\newcommand{\AddrPeking}{Center for High-Energy
Physics, Peking University, Beijing, 100871, P. R. China}
\newcommand{\AddrStauba}{Institute for Theoretical Physics (ITP), Karlsruhe Institute of Technology,
Engesserstra?e 7, D-76128 Karlsruhe, Germany}

\newcommand{\AddrStaubb}{Institute for Nuclear Physics (IKP), Karlsruhe Institute of Technology,
Hermann-von-Helmholtz-Platz 1, D-76344 Eggenstein-Leopoldshafen, Germany}

\begin{document}

\preprint{KA-TP-26-2017}

\title{Naturalness and a light $Z'$}

\author{Bin Zhu}
\affiliation{\AddrYantai}

\author{Florian Staub}
\affiliation{\AddrStauba}
\affiliation{\AddrStaubb}

\author{Ran Ding}
\affiliation{\AddrPeking}

\begin{abstract}
Models with a light, additional gauge boson are attractive extensions of the standard model. Often these models are only considered as effective low energy theory without any assumption about an UV completion. This leaves not only the hierarchy problem of the SM unsolved, but introduces a copy of it because of the new fundamental scalars responsible for breaking the new gauge group. A possible solution is to embed these models into a supersymmetric framework. However, this gives rise to an additional source of fine-tuning compared to the MSSM and poses the question how natural such a setup is. One might expect that the additional fine-tuning is huge, namely, $O(M^2_{\rm SUSY}/m^2_{Z'})$.
In this paper we point out that this is not necessarily the case. We show that it is possible to find a focus point behaviour also in the new sector in co-existence to the MSSM focus point. We call this 'Double Focus Point Supersymmetry'. Moreover, we stress the need for a proper inclusion of radiative corrections in the fine-tuning calculation: a tree-level estimate would lead to predictions for the tuning which can be wrong by many orders of magnitude. As showcase, we use the $U(1)_{B-L}$ extended MSSM and discuss possible consequence of the observed $^8\textrm{Be}$ anomaly. However, similar features are expected for other models with an extended gauge group which involve potentially large Yukawa-like interactions of the new scalars. 
\end{abstract}

\maketitle

\section{Introduction}

Supersymmetry (SUSY) is still one of the best candidates for physics beyond standard model (BSM), which provides an elegant solution to hierarchy problem, achieving gauge coupling unification at a high scale, and providing a dark matter candidate~\cite{Martin:1997ns}. However, the current null results from LHC direct SUSY searches together with the measured $125$ GeV Higgs mass have exacerbated the little hierarchy problem and put pressure on the Minimal Supersymmetric Standard Model (MSSM) as natural extension of the SM. The little hierarchy is usually encoded in the following equation:
\begin{align}
m_Z^2\sim -2m_{H_u}^2-2\mu^2~,
\label{eqn:litt}
\end{align}
in order to prevent large fine-tuning, one needs $m_Z^2\sim m_{H_u}^2\sim\mu^2$ at the SUSY scale what becomes more and more disfavoured. To quantify the resulting amount of tuning, one can use one of the common measures like the one 
of Barbieri-Giudice~\cite{Barbieri:1987fn}
\begin{equation}
\Delta^Z = \text{max} \left|\frac{\partial \ln m_Z^2}{\partial \ln \alpha} \right| 
\end{equation}
where $\alpha$ represents the fundamental parameters of the model. With this measure one finds that the tuning stemming from $\mu$ is at tree-level given by $\Delta_\mu \simeq 2 \mu^2/m_Z^2$. More recently, it was found that a more accurate prediction is actually $\Delta \simeq \mu^2/m_Z^2$ once loop corrections are taken into account \cite{Ross:2017kjc} \footnote{This connection between the supersymmetric $\mu$-term and the fine-tuning usually leads to the assumption that natural SUSY necessarily needs a light Higgsino mass. However, it was also pointed out that this strong conclusion can be avoided by introducing a non-holomorphic soft-term $\mu'$ \cite{Ross:2016pml,Ross:2017kjc}}. The other source of the fine-tuning is $m_{H_u}^2$ which shows a large sensitive on the radiative corrections from (s)tops and the gluino. Consequently, the fine-tuning in the MSSM has a significant correlation with the value of the SM-like Higgs mass which also depends strongly on these particles. The SUSY masses necessary to explain the measured mass of 125~GeV \cite{Aad:2012tfa,Chatrchyan:2012xdj} unavoidable introduce a sizeable amount of fine-tuning which is typically very large in most regions of the parameter space. This is especially the case in unified scenarios like the constrained MSSM (CMSSM) with only a small set of free parameters at the scale of grand unification (GUT). The most important exception is the focus point region where the fine-tuning has only a mild dependence on the stop masses~\cite{Feng:1999zg,Feng:2013pwa,Feng:2011aa,Feng:2012jfa,Feng:2000bp,
Horton:2009ed,Brummer:2012zc,Yanagida:2013ah,Agashe:1999ct,Draper:2013cka,Ding:2013pya,Kim:2013uxa}. 

The situation becomes more complicated when a new gauge group is introduced. We consider here the case of an additional Abelian group which gets broken by a pair of scalars $\eta_1$, $\eta_2$ with charges $\pm 1$ under this group. In this case, one should also consider the tadpole equation associated with these new states and calculate the corresponding amount of tuning needed to fulfil them. Under the assumption that the new superfields also receive their SUSY mass from a dimensionful term in the superpotential (called $\mu_\eta$ in the following), one finds at tree-level a very similar connection as eq.~(\ref{eqn:litt}) between the mass of the new gauge boson $m_{Z^{\prime }}^2$ and the SUSY parameters
\begin{align}
m_{Z^{\prime }}^2\sim -m_\eta^2-\mu_\eta^2~,
\label{eqn:zp}
\end{align}
If one assumes that the  $U(1)$ gauge coupling $g^\prime$ is of the same order as the weak coupling $g$, the current LHC bound for $m_{Z^\prime}$ is in the multi-TeV range~\cite{ATLAS-CONF-2016-045}. For such heavy $Z^\prime$, one does not need to worry too much about the fine-tuning induced by eq.~(\ref{eqn:zp}) \cite{Athron:2013ipa,Athron:2015tsa}. However, a very light $Z^\prime$ is still allowed for tiny $g^\prime$. In general, there are some motivations to consider such a light vector boson, e.g.
\begin{itemize}
\item Self-interacting dark matter (SIDM)~\cite{Davoudiasl:2010am,Tulin:2013teo,Boddy:2014yra} provides a  possibility to reconcile the tension between the small scale structure observations and the conventional cold DM (CDM) predictions  ~\cite{Flores:1994gz,BoylanKolchin:2011de,BoylanKolchin:2011dk}, see for instance Ref.~\cite{Tulin:2017ara} for a recent summary. 
One possible explanation for the self-interaction of DM is the exchange of light gauge bosons \cite{Mahoney:2017jqk}.

\item  An anomaly has been reported from isoscalar ${^8\textrm{Be}}^*(1^+) \to {^8\textrm{Be}}(0^+)+e^+e^-$ transitions: a bump in the opening angle distributions of $e^+e^-$ pairs with $6.8\sigma$ significance was observed~\cite{Krasznahorkay:2015iga}. This conflicts with the standard expectation, which predicts that the distribution of opening angles of $e^+e^-$ pairs should follow a smoothly downward curve.  In addition, in the related isovector ${^8\textrm{Be}}^{*\prime}(1^+)\to {^8\textrm{Be}}(0^+)+e^+e^-$ transition no excess is visible. If one takes this excess serious, it could be originate from a new light vector boson $X$ with the decay channel ${^8\textrm{Be}}^*(1^+) \to {^8\textrm{Be}}(0^+)+ X\to {^8\textrm{Be}}(0^+) + e^+e^-$~\cite{Krasznahorkay:2015iga,Feng:2016jff}. The necessary mass to fit the data is
    \begin{align}
    m_X= 16.7 \, \pm \,  0.35 (\text{stat})\, \pm \,  0.5(\text{sys}) \,\, \text{MeV}~.
    \end{align}
    Refs.~\cite{Feng:2016jff,Feng:2016ysn} have examined the case of a purely vector interaction with quarks. It has been shown that in order to be compatible with existing experimental constraints, the new vector boson should be protophobic~\cite{Feng:2016jff}, i.e., the coupling to proton is highly
    suppressed compared with the coupling to neutron. This proposal has stimulated many related works~\cite{Gu:2016ege,Ellwanger:2016wfe,Liang:2016ffe,Chen:2016kxw,
    Jia:2016uxs,Kitahara:2016zyb,Kahn:2016vjr,Kozaczuk:2016nma,Seto:2016pks}.
\end{itemize}
Such a light $Z'$ is expected to dominate the overall fine-tuning in the model if $m_\eta$ and $\mu_\eta$ are $O(M_{\rm SUSY})$ \footnote{We don't make in the following any assumption how the different fine-tunings shall be combined, but we discuss them separately. Often, the tunings are dominates by the different $\mu$-terms. In these cases the overall fine-tuning is given by $\Delta = \Delta^Z_\mu \times \Delta^{Z'}_{\mu_\eta}$. The situation is more complicated if the same parameter (like an universal scalar mass $m_0$) contributes significantly to the tuning in both sectors at the same time.}. \\

We want to study in the following for the first time explicitly the impact of a light $Z'$ on the fine-tuning in supersymmetric models. As example, we consider an $U(1)_{B-L}$ extension of the MSSM which was in the past mainly studied in the context of heavy $Z'$ masses \cite{Khalil:2007dr,CamargoMolina:2012hv,Basso:2012tr,Basso:2012gz,Staub:2015kfa,Belanger:2015cra,Un:2016hji,DelleRose:2017ukx}. In particular, we discuss the necessary conditions to reach a focus-point like behaviour in the new sector of the model. We show that the new focus point can co-exist with the well-know MSSM focus point resulting in a 'Double Focus Point' (DFP) scenario. We also discuss the radiative corrections to the fine-tuning. These corrections cause significant deviations in the actual fine-tuning prediction from the tree-level estimate $\Delta^{Z'}_{\mu_\eta} \sim \mu_\eta^2/m_{Z'}^2$. \\ 

The rest of paper is organized as follows: in section~\ref{sec:model} we present the details of the considered model. In section~\ref{sec:nbl} we show the analytical derivation of the DFP and calculate the dominant radiative corrections to the fine-tuning. In section~\ref{sec:fif} we perform a numerical study of the fine-tuning to validate our analytical results. Afterwards, we analyse the impact of the $^8\textrm{Be}$ anomaly. We conclude in section~\ref{sec:conclusion}.

\section{The $U(1)_{B-L}$ extended MSSM}
\label{sec:model}
In the simplest $U(1)_{B-L}$ extension of the MSSM, the chiral superfields are extended by a pair bileptons  ($\hat \eta_{1}, \hat \eta_2$) and three generations of  right-handed neutrino superfield $\hat \nu_{R_i}$. The complete particle contents and charge assignments are listed in table~\ref{tab:gauge} and \ref{tab:chiral}.
\begin{table}
\begin{center}
\begin{tabular}{|c|c|c|c|c|c|}
\hline \hline
Superfield & Spin \(\frac{1}{2}\) & Spin 1 & Gauge group & Coupling\\
\hline
$\hat{B}$ & \(\lambda_{\tilde{B}}\) & $B$ & $U(1)_Y$ & $g_1$ \\
$\hat{W}$ & \(\lambda_{\tilde{W}}\) & $W$ & $SU(2)_L$ & $g_2$ \\
$\hat{g}$ & \(\lambda_{\tilde{g}}\) & $g$ & $SU(3)_c$ & $g_3$ \\
$\hat{B}^\prime$ & \(\lambda_{\tilde{B}^\prime}\) & $B^\prime$ & $U(1)_{B-L}$ & $g_B$ \\
\hline \hline
\end{tabular}
\end{center}
\caption{
Vector superfields of the
BLSSM and corresponding gauge couplings.\label{tab:gauge}}
\end{table}
\begin{table}
\begin{center}
\begin{tabular}{|c|c|c|c|c|c|}
\hline \hline
Superfield & $N_G$ & $U(1)_Y\otimes\, SU(2)_L\otimes\, SU(3)_c\otimes\, U(1)_{B-L}$ \\
\hline
$\hat Q$  & 3 & $\frac{1}{6}\otimes\,{\bf 2}\otimes\,{\bf 3}\otimes\,\frac{1}{6}$ \\
$\hat U$ & 3 & $-\frac{2}{3}\otimes\,{\bf 1}\otimes\,{\bf \overline{3}}\otimes\,-\frac{1}{6}$ \\
$\hat D$ & 3 & $\frac{1}{3}\otimes\,{\bf 1}\otimes\,{\bf \overline{3}}\otimes\,-\frac{1}{6}$ \\
$\hat L$ & 3 & $-\frac{1}{2}\otimes\,{\bf 2}\otimes\,{\bf 1}\otimes\,-\frac{1}{2}$ \\
$\hat E$ & 3 & $1\otimes\,{\bf 1}\otimes\,{\bf 1}\otimes\,\frac{1}{2}$ \\
$\hat \nu_R$  & 3 & $0\otimes\,{\bf 1}\otimes\,{\bf 1}\otimes\,\frac{1}{2}$ \\
$\hat H_u$  & 1 & $\frac{1}{2}\otimes\,{\bf 2}\otimes\,{\bf 1}\otimes\,0$\\
$\hat H_d$  & 1 & $-\frac{1}{2}\otimes\,{\bf 2}\otimes\,{\bf 1}\otimes\,0$ \\
$\hat \eta_1$ & 1 & $0\otimes\,{\bf 1}\otimes\,{\bf 1}\otimes\,-1$ \\
$\hat \eta_2$ & 1 & $0\otimes\,{\bf 1}\otimes\,{\bf 1}\otimes\,1$ \\
\hline \hline
\end{tabular}
\end{center}
\caption{
Chiral superfields of the
BLSSM and their charges under $U(1)_Y\otimes\, SU(2)_L\otimes\, SU(3)_c\otimes\, U(1)_{B-L}$ gauge group.\label{tab:chiral}}
\end{table}
This model is known as the BLSSM and its superpotential is given by
\begin{align}
W = & Y^{ij}_u \hat U_i \hat Q_j \hat H_u\,- Y^{ij}_d \hat D_i \hat Q_j \hat H_d\,- Y^{ij}_e \hat E_i \hat L_j \hat H_d\, +\mu \hat H_u \hat H_d \nonumber\\
   & Y^{ij}_{\eta} \hat \nu_{Ri} \hat \eta_1 \hat \nu_{Rj}\,+Y^{ij}_\nu \hat L_i \hat H_u \hat \nu_{Rj}\, - {\mu_\eta} \hat \eta_1 \hat \eta_2~.
\label{eq:st}
\end{align}
Here $i,j$ denote family indices and all colour and isospin indices are suppressed. 
The soft-breaking terms are
\begin{align}
\mathcal{L}_{BLSSM} &= \mathcal{L}_{MSSM} - {M}_{B B^\prime} \lambda_{\tilde{B}} \lambda_{\tilde{B}^\prime} - \frac{1}{2} {M}_{B^\prime} \lambda_{\tilde{B}^\prime} \lambda_{\tilde{B}^\prime} \nonumber \\
& - m_1^2 |\eta_1|^2 - m_2^2 |\eta_2|^2  - {m_{\nu,ij}^{2}} (\tilde{\nu}_{Ri}^c)^* \tilde{\nu}_{Rj}^c   \nonumber \\
&- B_{\mu_\eta}\eta_1 \eta_2 + T^{ij}_{\nu}  H_u \tilde{\nu}_{Ri}^c \tilde{L}_j + T^{ij}_\eta \eta_1 \tilde{\nu}_{Ri}^c \tilde{\nu}_{Rj}^c
\end{align}
After Higgs states and bileptons receive vacuum expectation values (VEVs), the electroweak and $U(1)_{B-L}$ symmetry are broken to $U(1)_{em}$. After symmetry breaking, the complex scalars are parametrised by
\begin{eqnarray}
H_d^0 &=& \frac{1}{\sqrt{2}} \left(i \sigma_{d} + v_d  +  \phi_{d} \right)~,\,
H_u^0 = \frac{1}{\sqrt{2}} \left(i \sigma_{u} + v_u  +  \phi_{u} \right)~,\nonumber\\
\eta_1 &=& \frac{1}{\sqrt{2}} \left(i \sigma_1 + v_1 +  \phi_1 \right)~, \, \eta_2 = \frac{1}{\sqrt{2}} \left(i \sigma_2 + v_2  +  \phi_2 \right)~.
\end{eqnarray}
Following the MSSM definition $\tan\beta = v_u/v_d$, we denote the ratio of the two bilepton VEVs as $\tan\beta' = v_1/v_2$.

The particle content of the BLSSM gives rise to gauge-kinetic mixing even if it is absent at a given scale. This introduces two additional 
gauge couplings $g_{YB}\equiv g_{Y\,B-L}$ and $g_{BY}\equiv g_{B-L \, Y}$, i.e. the general form of the covariant derivatives is
\begin{equation}
D_\mu \phi = \left(\partial_\mu - i \sum_{i,j} Q_\phi^i g_{ij}   V^\mu_j \right)\phi \hspace{1cm} i,j = Y,B-L
\end{equation}
Here, $Q_\phi^i$ is the $U(1)$ charge of the particle $\phi$ under the gauge group $U(1)_i$ ($i=Y,B-L$).  \\
However, as long as the two Abelian gauge groups are unbroken, we are allowed to make a change of basis. This freedom is used go to a basis where electroweak  precision data is respected in a simple way:  by choosing a triangle form of the gauge coupling matrix, the bilepton contributions to the $Z$ mass vanish:
\begin{equation}
\left(\begin{array}{cc} g_{YY} & g_{YB} \\ g_{BY} & g_{BB}   \end{array} \right) \to  \left(\begin{array}{cc} g_1 & \tilde{g} \\ 0 & g_B \end{array} \right)~,
\end{equation}
and the gauge couplings are related by  \cite{Chankowski:2006jk}:
\begin{eqnarray}
\label{eq:Triangle1}
g_1 &=& \frac{g_{YY} g_{BB} - g_{YB} g_{BY}}{\sqrt{g_{BB}^2 + g_{BY}^2}}~, \nonumber\\
\tilde{g} &=& \frac{g_{YB} g_{BB} + g_{BY} g_{YY}}{\sqrt{g_{BB}^2 + g_{BY}^2}}~, \nonumber\\
g_B &=& \sqrt{g_{BB}^2 + g_{BY}^2}~.
\label{eq:triangle}
\end{eqnarray}
In addition, After electroweak and $U(1)_{B-L}$ breaking, the gauge-kinetic mixing further induces a mixing between the neutral SUSY particles from the MSSM and from the new sector, i.e. 
there are seven neutralinos in this model. In the gauge sector, the three neutral gauge bosons $B$, $W^3$ and $B^\prime$ are rotated to the mass eigenstates $\gamma$, $Z$, $Z^\prime$ via:
\begin{align}
\left(\begin{array}{c}
B\\
W\\
{B^\prime}\end{array} \right)
 = R \left(\theta,~\theta^\prime\right)
\left(\begin{array}{c}
\gamma\\
Z\\
{Z^\prime}\end{array} \right)~,
\end{align}
where the rotation matrix $R (\theta,~\theta^\prime)$ depends on two angles $\theta$ and $\theta^\prime$ with following expression
\begin{align}
\label{eq:rotZZp}
\left(
\begin{array}{ccc}
\cos\theta_W & -\cos{\theta^\prime}_W \sin\theta_W &  \sin\theta_W \sin{\theta^\prime}_W \\
\sin\theta_W &  \cos\theta_W \cos{\theta^\prime}_W & -\cos\theta_W \sin{\theta^\prime}_W \\
 0 & \sin{\theta^\prime}_W  & \cos{\theta^\prime}_W \end{array}
\right)~.
\end{align}
The entire mixing between the $U(1)_{B-L}$ and the SM gauge depends on mixing angle  $\theta^\prime$, which can be approximately expressed as~\cite{Basso:2010jm}
\begin{equation}
\label{eq:ThetaWP}
\tan 2 {\theta'}_W \simeq \frac{2 \tilde{g} \sqrt{g_1^2 + g_2^2}}{\tilde{g}^2 + 16 \left(v_\eta/v\right)^2 g_B^2 -g_2^2 - g_1^2}~.
\end{equation}
with $v=\sqrt{v_u^2+v_d^2}$ and $v_\eta=\sqrt{v_1^2 + v_2^2}$.

\section{Focus point behaviour and loop corrected fine-tuning in the BLSSM}
\label{sec:nbl}
After setting up the model, we begin to investigate the focus point property. Compared to the MSSM, the tadpole equations become more complicated and we just present the most relevant ones:
\begin{align}
m_{H_d}^2 &= \frac18 D_H + \frac{1}{\tan\beta} B_\mu - \mu^2 \label{eqn:tad1}\\
m_{H_u}^2 &= -\frac18 D_H + {\tan\beta} B_\mu - \mu^2\\
m_{\eta_1}^2 &= \frac{1}{4}\,D_\eta+  \frac{1}{\tan\beta'} B_{\mu_\eta} -\mu_\eta^2~, \\             
m_{\eta_2}^2 &= -\frac{1}{4}\,D_\eta+  \tan\beta' B_{\mu_\eta} -\mu_\eta^2~.
   \label{eqn:tad2}
\end{align}
with $D_H = (g_1^2 + \tilde{g}^2 + g_2^2) v^2 \cos2\beta + 2 \tilde{g} g_B v_\eta^2 \cos2\beta' $ and $D_\eta = \big(v^2 \cos2\beta g_B \tilde{g}+2 v_\eta^2 \cos2\beta' g_B^2  \big)$.
It is easy to see that for small kinetic mixing coupling $\tilde{g}$, the MSSM sector and $U(1)_{B-L}$ sector are decoupled. Therefore, the little hierarchy problems for two sectors can be treated separately. Nevertheless, there will be a non-trivial link between the fine-tuning of both sectors because of the relations of soft-breaking terms at the GUT scale. 
Under this assumption eq.~(\ref{eqn:tad1}) reduces to the standard MSSM tadpole equation and the little hierarchy problem can be handled as usual. In the $B-L$ sector we have to distinguish
the cases of small $\tan\beta'$ ($\simeq 1$) and large $\tan\beta'$ ($\gg 1$) which we discuss separately in the following. 

\subsection{Small $\tan\beta'$}
In analogy to the MSSM, the tadpole equation of the $B-L$ sector,  eq.~(\ref{eqn:tad2}), can be simplified for $\tan\beta' \simeq 1$ to
\begin{align}
m^2_{Z^\prime}&=-m_{\eta_2}^2-\mu_\eta^2~,
\label{eqn:tad4}
\end{align}
Here, we used $m^2_{Z^\prime}=g_B^2 v_\eta^2$. The important question is now, if it possible to obtain naturally values for $m_\eta$ and $\mu_\eta$ which are significantly
smaller than the ordinary SUSY parameters during the renormalisation group equation (RGE) evolution. Or, to phrase it differently: is it possible to find a focus point behaviour? To answer this question, one needs to check the 
running of $m^2_{\eta_2}$. The one-loop beta function is given by
\begin{align}
\frac{d m_{\eta_2}^2}{dt}=\frac{g_B^2}{16\pi^2}(-12 M_{B B^\prime}^2+M_{B^\prime}^2)~.
\label{eqn:beta1}
\end{align}
where $M_{B B^\prime}$ is a soft-breaking term mixing the two gaugino fields. Due to the absence of any Yukawa interaction in the running, it is impossible to obtain a focus point behaviour in this scenario. $m_{\eta_2}$ will always increase during the evaluation from the GUT to the SUSY scale.  Thus, in this simplest realisation, it is not possible to obtain a DFP for small $\tan\beta'$. However, this would become possible in the singlet extension of the model (N-BLSSM), with an additional superpotential term 
\begin{align}
\delta W=\lambda\, \hat S\,\hat \eta_1\, \hat \eta_2
\end{align}
$\lambda$ will give new contributions to the running of $m_{\eta_2}^2$. We discuss this briefly in appendix \ref{app:nblssm}.

\subsection{Large $\tan\beta'$}
We turn now to the case of large $\tan\beta'$ which is usually not considered in the case of heavy $Z^{\prime}$ masses. The reason is that the new $D$-term contributions to the sfermion masses result in tachyonic states once $\tan\beta'$ is too large. However, this is not the case for a light $Z^{\prime}$. The corresponding tadpole equation in this case is given by
\begin{align}
m^2_{Z^\prime}&=-m_{\eta_1}^2-\mu_\eta^2~.
\label{eqn:tad5}
\end{align}
The one-loop running of  $m_{\eta_1}^2$ is given by
\begin{align}
\frac{d m_{\eta_1}^2}{dt}=\frac{y_{\eta}^2}{8\pi^2}(4m_{\eta_1}^2+4m_{\nu_R}^2+4 A_\eta^2)~.
\label{eqn:beta2}
\end{align}
Here the soft trilinear term $A_\eta=T^{ij}_\eta/Y^{ij}_\eta$. We assumed here that $y_\eta$ as well as $m_{\nu_R}^2$ are diagonal and degenerated. We defined $y_{\eta} \equiv Y^{ii}_\eta$. We can now investigate the necessary conditions to obtain a focus point in the running. Before we proceed, the following comments are at place
\begin{itemize}
 \item We assume the beta functions of $m_{H_u}^2$ and $m_{\eta_1}^2$ to be independent from each other. This is justified as long as gauge kinetic mixing is small. 
\item Since the beta function of $m_{H_u}^2$ depends on $m_{q}^2$, $m_{u}^2$ and $A_t^2$, we must solve the coupled system of equations. In the $B-L$ sector we need to consider $m_{\eta_1}$,
$A_{\eta}$ and $m_{\nu_R}$. 
\item In contrast to the MSSM sector, where the top Yukawa coupling is fixed by experiment, we can treat $y_\eta$ as free parameter. If we demand perturbativity up to $M_{\rm GUT})$, the maximal allowed value for $y_\eta$ is about 0.42. 
\end{itemize}
The relevant one-loop beta functions can be written into matrix form. The MSSM part is given by

\begin{align}
\frac{d}{dt}\left[
              \begin{array}{c}
                m_{H_u}^2 \\
                m_{u}^2 \\
                m_{q}^2 \\
                A_t^2 \\
              \end{array}
            \right]
            =\frac{y_t^2}{8\pi^2}
            \left[
                    \begin{array}{cccc}
                      3 & 3 & 3 & 3 \\
                      2 & 2 & 2 & 2 \\
                      1 & 1 & 1 & 1 \\
                      0 & 0 & 0 & 12 \\
                    \end{array}
            \right]
            \left[
              \begin{array}{c}
                m_{H_u}^2 \\
                m_{u}^2 \\
                m_{q}^2 \\
                A_t^2 \\
              \end{array}
            \right]~,
\end{align}
and the BLSSM one by
\begin{align}
\frac{d}{dt}\left[
              \begin{array}{c}
                m_{\eta_1}^2 \\
                m_{\nu_R}^2 \\
                A_{x}^2
              \end{array}
            \right]
            =\frac{y_{\eta}^2}{8\pi^2}
            \left[
                    \begin{array}{cccc}
                      6 & 12 & 6 \\
                      4 & 8 & 4 \\
                      0 & 0 & 28 \\
                    \end{array}
            \right]
            \left[
              \begin{array}{c}
                m_{\eta_1}^2 \\
                m_{\nu_R}^2 \\
                A_{\eta}^2 \\
              \end{array}
            \right]
            \label{eqn:beta3}
\end{align}
The coupled beta functions can be solved in terms of eigenvectors and eigenvalues. We obtain
\begin{align}
\left[
              \begin{array}{c}
                m_{H_u}^2 \\
                m_{u}^2 \\
                m_{q}^2 \\
                A_t^2 \\
              \end{array}
            \right]
            &=
            \kappa_{12}\left[
              \begin{array}{c}
                3 \\
                2 \\
                1 \\
                6 \\
              \end{array}
            \right]\operatorname{e}^{12I[t]}
           +
            \kappa_{6}\left[
              \begin{array}{c}
                3 \\
                2 \\
                1 \\
                0 \\
              \end{array}
            \right]\operatorname{e}^{6I[t]}\nonumber\\
           &+
           \kappa_{0}\left[
              \begin{array}{c}
                -1 \\
                0 \\
                1 \\
                0 \\
              \end{array}
            \right]
            +
            \kappa_{0}^{\prime}\left[
              \begin{array}{c}
                -1 \\
                1 \\
                0 \\
                0 \\
              \end{array}
            \right]~,
\end{align}

as well as

\begin{align}
\left[
              \begin{array}{c}
                m_{\eta_1}^2 \\
                m_{\nu_R}^2 \\
                A_{\eta}^2 \\
              \end{array}
            \right]
            &=
            \epsilon_{28}\left[
              \begin{array}{c}
                3 \\
                2 \\
                7 \\
              \end{array}
            \right]\operatorname{e}^{28K[t]}
           +
            \epsilon_{14}\left[
              \begin{array}{c}
                3 \\
                2 \\
                0 \\
              \end{array}
            \right]\operatorname{e}^{14K[t]}\nonumber\\
           &+
           \epsilon_{0}\left[
              \begin{array}{c}
                -2 \\
                1\\
                0 \\
              \end{array}
            \right]
\end{align}
with so far arbitrary coefficients $\kappa_i$ and $\epsilon_i$. 
The functions $I$ and $K$ are defined in Appendix~\ref{sec:app}. Approximate values 
are $\operatorname{e}^{14K[t]} \simeq 1/10$ and 
$\operatorname{e}^{6I[t]} \simeq 1/3$. This results in 

\begin{align}
\left[\begin{array}{c}
                m_{H_u}^2[Q_{\text{GUT}}] \\
                m_{q}^2[Q_{\text{GUT}}] \\
                m_{u}^2[Q_{\text{GUT}}] \\
                A_{t}^2[Q_{\text{GUT}}]
              \end{array}\right]
            &  =\left[\begin{array}{c}
                m_{0}^2 \\
                m_{0}^2+\kappa _0^{\prime}-\frac{2 \kappa
   _{12}}{3} \\
                m_{0}^2-\kappa _0^{\prime}-\frac{4 \kappa
   _{12}}{3} \\
                6\kappa_{12}
              \end{array}\right]\nonumber\\
&\rightarrow\nonumber\\
\left[\begin{array}{c}
                m_{H_u}^2[Q_{\text{SUSY}}] \\
                m_{q}^2[Q_{\text{SUSY}}] \\
                m_{u}^2[Q_{\text{SUSY}}] \\
                A_{t}^2[Q_{\text{SUSY}}]
              \end{array}\right]
              &=\left[\begin{array}{c}
                0 \\
                \frac{m_{0}^2}{3}+\kappa _0^{\prime}-\frac{2 \kappa
   _{12}}{5} \\
                \frac{2m_{0}^2}{3}+\kappa _0^{\prime}-\frac{4 \kappa
   _{12}}{5} \\
                \frac{2}{3}\kappa_{12}
              \end{array}\right]~,
\label{eqn:focus1}
              \end{align}
              
and              

\begin{align}
\left[\begin{array}{c}
                m_{\eta_1}^2[Q_{\text{GUT}}] \\
                m_{\nu_R}^2[Q_{\text{GUT}}] \\
                A_{\eta}^2[Q_{\text{GUT}}
              \end{array}\right]
 &  =\left[\begin{array}{c}
                m_{0}^2 \\
                \frac{43}{54}m_0^2-\frac{7}{20}\epsilon_{28} \\
                7\epsilon_{28}
              \end{array}\right]\nonumber\\
 &\rightarrow\nonumber\\
    \left[\begin{array}{c}
                m_{\eta_2}^2[Q_{\text{SUSY}}] \\
                m_{\nu_R}^2[Q_{\text{SUSY}}] \\
                A_{\eta}^2[Q_{\text{SUSY}}]
              \end{array}\right]
             & =\left[\begin{array}{c}
                0 \\
               \frac{7}{54}m_0^2-\frac{7}{20}\epsilon_{28} \\
                \frac{7}{100}\epsilon_{28}
              \end{array}\right]
\label{eqn:focus3}
          \end{align}

Eq.~(\ref{eqn:focus2}) and~(\ref{eqn:focus3}) show that $m_{H_u}^2$ and $m_{\eta_1}^2$ evolve to zero at low scale no matter what value we take for $m_0$ which means that the weak scale and $U_{B-L}$ breaking scale are insensitive to variation of fundamental parameters. As a consequence, we obtain DFP SUSY even for several TeV sfermions which are induced by large $m_0$. A few more comments:
\begin{enumerate}
  \item Besides $m_0$, there are three parameters: $\kappa_{12}$, $\kappa_0^{\prime}$, and $\epsilon_{28}$. Here $\kappa_{12}$ and $\epsilon_{28}$ represent large A-term generated by gravity mediation. 
  \item The parameters $\kappa_0^{\prime}$ and $\kappa_{14}$ give deviation from the soft masses are predicted by minimal supergravity. The source of the deviation could  be for instance hybrid anomaly or gauge mediation.
\item The parameters $\kappa_{12}$,$\kappa_0$ and $\epsilon_{28}$ are all dimensionful. They can be related to $m_0$ via dimensionless parameters $x$,$y$ and $z$ respectively. This turns out to be helpful for the numerical calculation. 
\end{enumerate}

\subsection{Radiative corrections to the fine-tuning}
\begin{figure} [tb]
\begin{center}
\includegraphics[width=0.40\textwidth]{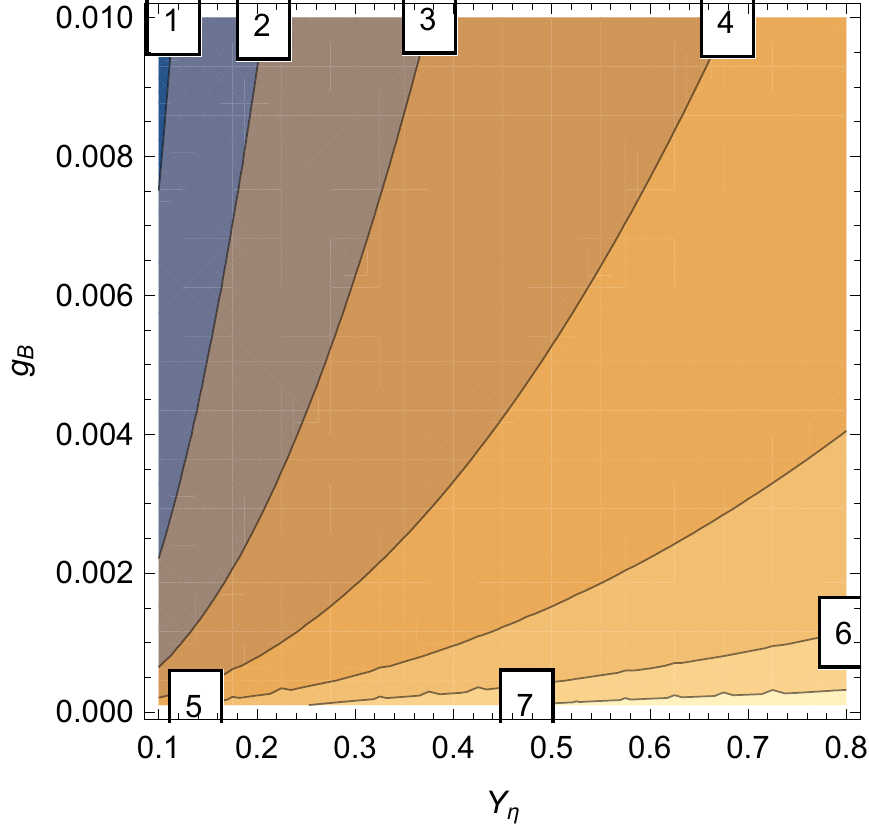}
\end{center}
\caption{The improvement in the fine-tuning as function of $g_B$ and $y_\eta$ when including radiative corrections to the tadpole equations. The lines correspond to contours of constant 
$\log_{10} \left({\Delta_{\mu_\eta}^{\rm Tree}}/{\Delta_{\mu_\eta}^{\rm Loop}}\right)$. We used here $m_{Z^{\prime}}=100$~MeV and $M_{\rm SUSY}=2.5$~TeV.}
\label{fig:oneloop}
\end{figure}

Up to now, we have shown that it is possible to find naturally parameter regions in which $|\mu|$ and $|\mu_\eta|$ at the same time are significantly smaller than the ordinary SUSY scale. 
Nevertheless, $\mu_\eta$ is still expected to be of the same size as $\mu$, i.e. O(100~GeV). In that case, the tree-level estimate for the fine-tuning would be $\Delta_{\mu_\eta} \sim {\mu^2_\eta}/{m_Z^{\prime,2}}$ which predicts values for the fine-tuning above $10^6$ for $\mu_\eta > 100$~GeV and $m_Z^{\prime} < 100$~MeV. In that case, the overall fine-tuning would be completely dominated by the new sector and all the considerations about DFP wouldn't have been necessary at all. However, it has rather recently been pointed out in Ref.~\cite{Ross:2017kjc} that loop corrections to the fine-tuning are very important in the MSSM. The same kind of corrections is even more important here. The starting point of the discussion is the one-loop corrected tadpole equation which is in general given by
\begin{equation}
0 = m^2_{\eta_1} v_\eta + \mu^2_\eta v_\eta + \frac12 g_B^2 v_\eta^3 + \delta t_\eta
\end{equation}
with 
\begin{equation}
\delta t_\eta = \frac{\partial V^{(1)}}{\partial v_\eta}
\end{equation}
Here, $V^{(1)}$ is the one-loop effective potential which can be calculated as usual as \cite{Coleman:1973jx}
\begin{equation}
V^{(1)} = \frac{1}{16\pi^2} \sum_{i}^{\text{all fields}} r_i s_i C_i m_i^4 \left(\log\frac{m_i^2}{Q^2} - c_i \right)
\end{equation}
with $r_i = 1$ for real bosons, otherwise 2; $C_i$ is a colour factor; $\{s_i,c_i\}=\{-\frac{1}{2},\frac{3}{2}\}$ for fermions, $\{\frac{1}{4},\frac{3}{2}\}$ for scalars 
and $\{\frac{3}{4},\frac{5}{6}\}$ for vector bosons. In the case of large $y_\eta$ as needed for DFP, the dominant contributions are due to right (s)neutrinos. \\
In general,  $\delta t_\eta$ can be parametrised as
\begin{equation}
\delta t_\eta =  v_\eta \delta_1 + v_\eta^2 \delta_2 + v_\eta^3 \delta_3
\end{equation}
Because of symmetry reasons, $\delta_2$ always vanishes. Therefore, the fine-tuning with respect to $\mu_\eta$ becomes 
\begin{equation}
\label{eq:DZp}
\Delta_{\mu_\eta} = \frac{\mu^2_\eta}{(g_B^2 + 2 \delta_3)v_\eta^2} 
\end{equation}
Thus, for a reliable calculation of the fine-tuning, the knowledge of $\delta_3$ is crucial. One fiends for three generations of degenerated right sneutrinos with masses $O(M_{\rm SUSY})$ that $\delta_3$ at the SUSY scale is given by
\begin{equation}
\label{eq:d3}
\delta_3 = -\frac{3 y_\eta^4}{2 \pi^2} \log\frac{2 v_\eta^2 y_\eta^2}{M^2_{\rm SUSY}}
\end{equation}
From this one obtains an improvement in the fine-tuning at the loop level of
\begin{equation}
\label{eq:gainfactor}
\frac{\Delta_{\mu_\eta}^{\rm Tree}}{\Delta_{\mu_\eta}^{\rm Loop}} \sim \frac{3 y_\eta^4}{g_B^2 \pi^2} \log\frac{g_B^2 M_{\rm SUSY}^2 + 2 m_Z^{\prime,2} y_\eta^2}{2 m_Z^{\prime,2} y_\eta^2}
\end{equation}
The gain in the fine-tuning as function of $y_\eta$ and $g_B$ for $m_{Z^{\prime}}=100$~MeV is shown in Figure \ref{fig:oneloop}.
One might be surprised by these huge changes in the fine-tuning due to the radiative corrections. However, the same radiative corrections push also the mass of the light $B-L$ scalar, which is at tree-level $O(m_{Z^{\prime}})$, into the multi GeV range. Thus, the fine-tuning regarding the new vector boson mass ($\partial \ln m^2_{Z^{\prime}}/\partial \ln {\mu_\eta}$) becomes comparable to the fine-tuning regarding the pole mass of the scalars ($\partial \ln \log m^2_{h_\eta}/\partial \ln {\mu_\eta}$). Moreover, since higher order corrections don't modify the general form of the radiatively corrected tadpole equation but are only corrections to the coefficients $c_i$, similar huge changes in the fine-tuning won't occur by going to a higher loop level. Therefore, the numerical most important effects are already caught by the one-loop corrections \footnote{It has been discussed in the context of other models that the fine-tuning measure with respect to $m_Z$ can clearly deviate from a measure with respect to $m_h$ \cite{Kaminska:2014wia}. It would be interesting to see if a proper inclusion of loop corrections in the fine-tuning calculation reduces this discrepancy.}.

\section{Numerical Results}
\label{sec:fif}

\subsection{General results}
\begin{figure} [tb]
\begin{center}
\includegraphics[width=0.5\textwidth]{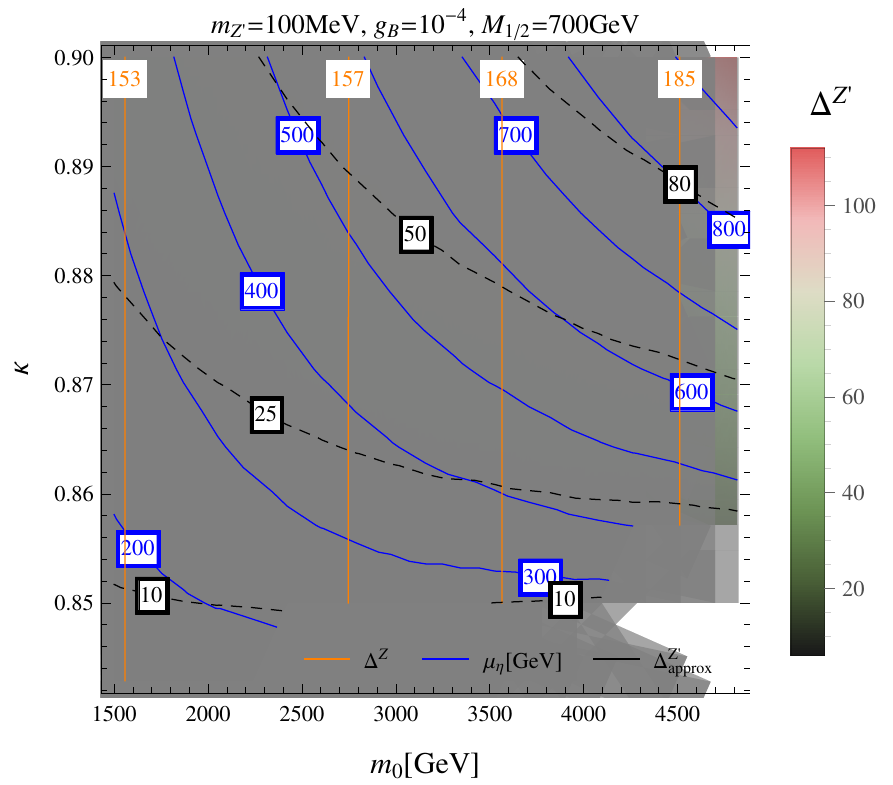} \\[5mm]
\includegraphics[width=0.5\textwidth]{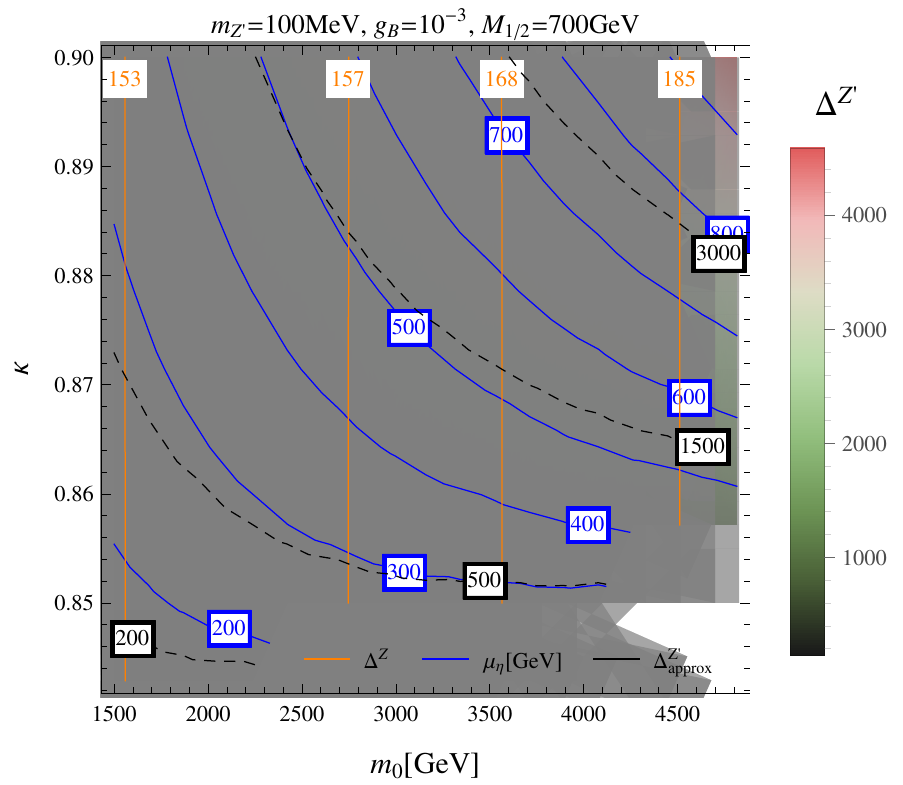}
\end{center}
\caption{The fine-tuning in the $(m_0,\kappa)$ plane for $m_{Z'}=100$~MeV and $g_B=10^{-4}$ (top) or $g_B=10^{-3}$ (bottom). We set $M_{1/2}=700$~GeV, $\tan\beta=20$, $\tan\beta'=10$, $Y_\eta=0.4$, $s_y=-0.1$, $s_z=s_x=0$, $\tilde{g}(Q_{\rm GUT})=0$. In the gray shaded areas the SM-like Higgs doesn't fall into the desired range of [122,128]~GeV. The orange line is the electroweak fine-tuning with respect $m_Z$ (which is dominated by $M_{1/2}$), the blue lines show $\mu_\eta$ (in GeV) and the dashed black line give the calculated fine-tuning using the approximations eqs.~(\ref{eq:DZp}) and (\ref{eq:d3}). }
\label{fig:loopFT}
\end{figure}
We are going to compare the analytical results of the last section with a fully numerical calculation to check the validity of our results.
For this purpose, we have implemented the considered model into the Mathematica package {\tt SARAH} \cite{Staub:2008uz,Staub:2009bi,Staub:2010jh,Staub:2012pb,Staub:2013tta}\footnote{We could
use the already existing implementation of the BLSSM which we had slightly to modify: since the eigenstates are mass ordered, the definition of the neutral gauge boson mixing 
was changes from $\gamma-Z-Z'$ to $\gamma-Z'-Z$ and the boundary conditions needed to be adjusted.}. We used this implementation to generate a spectrum generator for the model 
based on {\tt SPheno}~\cite{Porod:2003um,Porod:2011nf}. {\tt SPheno} solves numerically the full two-loop RGEs, calculates the mass spectrum 
at the full one-loop level  and includes all important two-loop correction to the neutral scalar masses \cite{Goodsell:2014bna,Goodsell:2015ira,Braathen:2017izn}. In order to keep the uncertainty of the Higgs mass to a low level also in the presence of very heavy SUSY scales, {\tt SPheno} provides an effective calculation within the SM \cite{Staub:2017jnp} where all SUSY effects are absorbed into $\lambda$ via a pole mass matching of the Higgs masses at the SUSY scale \cite{Athron:2016fuq}.  
Also a routine to obtain the electroweak fine-tuning is available out-of-the-box. This routine has been extended to calculate also the fine-tuning with respect to $Z'$ as
\begin{equation}
\Delta^{Z'} =  \text{max} \left|\frac{\partial \ln m_{Z'}^2}{\partial \ln \alpha} \right| 
\end{equation}
with $\alpha=\{m_0,M_{1/2},\mu'\}$. To obtain the loop corrected fine-tuning, we calculate the one-loop corrected tadpole equations with a diagrammatic approach and solve them numerically with respect to all four VEVs using a broydn routine. The VEVs obtained in that way are then used to calculate the dependence of the vector boson masses on a finite variation of the parameters at the GUT scale. We checked that the fine-tuning obtained in that way is independent of the chosen size of the finite variation if a sufficiently small value of $10^{-6}$ is chosen. All parameter scans have been carried out using the package~{\tt SSP}~\cite{Staub:2011dp}.\\  
One important question is how accurate the coefficient of $\frac{43}{54}\simeq 0.8$ for the soft-term of the right sneutrino soft-term is at the GUT scale. So far, we discussed only the one-loop running, but haven't considered the impact of two-loop RGEs. In order to check this, we use a more general parametrisation 
\begin{equation}
m_{\nu_R}^2[Q_{\text{SUSY}}] = \kappa\, m_0^2 \, {\bf 1}
\end{equation}
and treat $\kappa$ is free parameter. The other GUT conditions are
\begin{eqnarray}
& m_{H_u}^2 = m_{H_d}^2 = m_{\eta_1}^2 = m_{\eta_2}^2 \equiv m_0^2 & \nonumber \\
& m_l^2 = m_d^2 = m_e^2 \equiv m_0^2 {\bf 1} & \nonumber \\
& m_{q,11}^2 = m_{q,22}^2 = m_{u,11}^2 = m_{u,22}^2 \equiv m_0^2 & \nonumber \\
& m_{q,33} \equiv (1-x) m_0^2 \hspace{1cm} m_{u,33} \equiv (1+x-3y)m_0^2 & \nonumber \\
&M_1 = M_2 = M_3 = M_B \equiv M_{1/2}& \nonumber \\
&T_i \equiv 3 y m_0 Y_i \, (i=e,d,u) \hspace{0.5cm} T_\eta  \equiv \sqrt{7} z Y_\eta &
\end{eqnarray}
The calculated fine-tuning in the $(m_0,\kappa)$ plane for a $Z'$ mass of 100~MeV and different values of $g_B$ is shown in Figure~\ref{fig:loopFT}. We find that the 
numerical calculation confirms our analytical results to a large extent: the focus point behaviour can be observed for $\kappa$ values below 0.86 even for large $m_0$. Also the numerically calculated fine-tuning agrees with our analytical approximation. It is also found that the fine-tuning has a strong dependence on the 
value of $g_B$ for fixed $m_{Z'}$ as expected from eq.~(\ref{eq:gainfactor}). These results confirm that a light $Z'$ mass in supersymmetric models doesn't lead unavoidably to a huge fine-tuning as a one might expect. However, two conditions must be fulfilled: (i) the presence of a focus point, (ii) a Yukawa-like coupling to the scalars which give mass to the $Z'$ which is much bigger than the corresponding gauge coupling. A more detailed parameter scan of the model is beyond the scope of this paper and we will only discuss one more aspect: what is the size of the expected fine-tuning if the $^8\text{Be}$ should be explained within this model.

\subsection{The $^8\text{Be}$ anomaly}
\begin{figure} [tb]
\begin{center}
\includegraphics[width=0.40\textwidth]{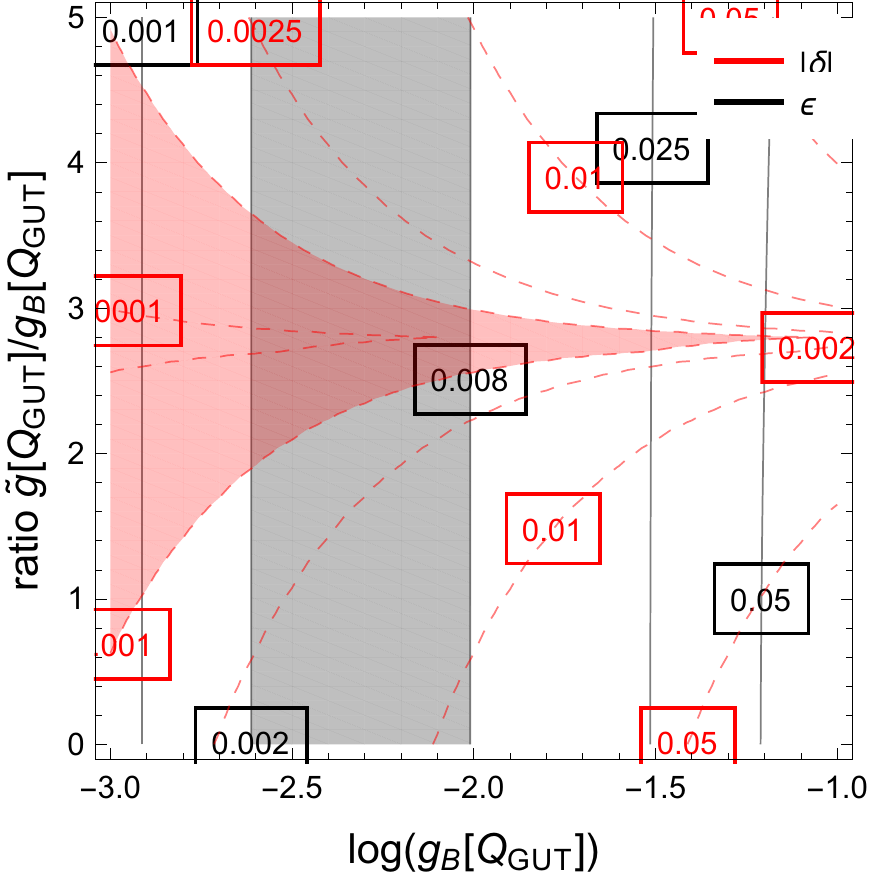}
\end{center}
\caption{Dependence of the ratio between $g_0$ and $g_d$ on $g_d$ at GUT scale.  Regions shaded in grey are the allowed region that explains Be anomaly.
The red dashed contour corresponds the abstract value of $\delta$ and black corresponds to  $\epsilon$.}
\label{fig:ratio}
\end{figure}

It has been shown in Ref.~\cite{Feng:2016ysn} that the $^8\text{Be}$ could be explained by a B-L gauge boson with a mass of about 17~MeV. However, strong constraints on the new couplings exist, especially on the one induced by gauge kinetic mixing. In general, the couplings to the SM fermions for the $B-L$ vector boson are given by
\begin{align}
g_u&=\frac{1}{3}g_{B}+\frac{2}{3}\tilde{g}~, \hspace{1cm} g_{\nu}=-g_{B} \nonumber\\
g_d&=\frac{1}{3}g_{B}-\frac{1}{3}\tilde{g}~, \hspace{1cm} g_{e}=-g_{B}-\tilde{g}~.
\label{eqn:cons}
\end{align}
which can be rewritten to 
\begin{align}
g_u&=-\frac{1}{3}\epsilon+\frac{2}{3}\delta~, \hspace{1cm} g_{\nu}=-\epsilon \nonumber\\
g_d&=+\frac{2}{3}\epsilon-\frac{1}{3}\delta~, \hspace{1cm}  g_{e}=-\delta~.
\label{eqn:cons2}
\end{align}
by defining $g_B=\epsilon$ and $\delta=-g_{B}+\tilde{g}$. The authors in Ref.~\cite{Feng:2016ysn} showed that, if the signal is real, $\delta$ and $\epsilon$ must fulfil
\begin{align}
& 0.002<|\epsilon|<0.008  ~, \hspace{1cm}
|\delta|<0.001
\end{align}
These constraints for $\epsilon$ and $\delta$ can be translated into constraints on the gauge couplings at the GUT scale. The preferred parameter ranges for $g_B({\rm GUT})$ and $\tilde{g}({\rm GUT})$ are shown in Fig.~\ref{fig:ratio}. 
%
Here, we used the two-loop SUSY RGEs with gauge kinetic mixing as calculated by  {\tt SARAH} based on the generic results of Ref.~\cite{Fonseca:2011vn}. One can see that $g_B$ must be smaller than 0.01 at the GUT scale. In contrast, $\tilde{g}$ must be bigger by a factor 2 to 4. This is already an unexpected hierarchy in the gauge couplings which might be hard to realize in a full model like $E_6 \times E_6$ which gets broken to $G_{SM} \times U(1)_{B-L}$ \cite{Buchmuller:2006ik,Ambroso:2009sc,Ambroso:2010pe}. Nevertheless, we take these parameter ranges as given. Since $g_B$ and $\tilde{g}$ are still sufficiently small to have only a weak link between the MSSM and BLSSM section, the discussion in the last section about the DFP behaviour remains fully valid. However, as expected from the analytical discussion and from the general results in the last subsection one must expect that the fine-tuning is rather large: for the values of $g_B$ necessary to explain the anomaly one expects only an improvement of the fine-tuning at the loop level by a few orders of magnitude, i.e. the tiny $Z'$ mass still causes an enormous fine-tuning even when the radiative corrections are included. \\
We performed a random scan for this model fixing $m_{Z'}$ as well as the new gauge couplings. The features of an representative parameter point which is in agreement with the Higgs 
mass measurement and which could explain the $^8\text{Be}$ anomaly is summarised in Table~\ref{tab:BP}. The fine-tuning in the MSSM sector is $O(170)$ and dominated by $M_{1/2}$ because of the gluino mass limit. Thus, to further improve this tuning, it would be necessary to give up gaugino unification to find a 'gaugino focus point' \cite{Horton:2009ed,Kaminska:2013mya}. In the $B-L$ sector the tuning with respect to $m_0$ and $\mu_\eta$ are of similar size and above 3000. This is not terribly good, but better than one might have expected. One possibility to further improve the tuning would be to assume a cut-off scale well below $10^{16}$~GeV: in that case the tuning with respect to $m_0$ becomes smaller because of the shorter running, and the tuning with respect to $\mu_\eta$ could be further reduce by larger $y_\eta$ because of the relaxed perturbativity condition.    \\

\begin{table}[tb]
\begin{tabular}{>{\centering\arraybackslash}m{1.25cm} >{\centering\arraybackslash}m{1.25cm} >{\centering\arraybackslash}m{1.25cm} >{\centering\arraybackslash}m{1.25cm} >{\centering\arraybackslash}m{1.25cm} >{\centering\arraybackslash}m{1.25cm} @{}m{0pt}@{} }
\hline 
\hline 
\multicolumn{7}{c}{\bf Input} \\
\hline 
$m_0$  & $M_{1/2}$ & $x$ & $y$ & $z$ & $\kappa$ &\\
4600  & 680 & 0.22 & -0.45 & 0.1 & 0.83 &\\[3mm]
$\tan\beta$ & $\tan\beta'$ & $y_\eta$ &  $m_{Z'} $ & $g_B$ & $\tilde{g}$  &\\
18.5 & 10 &0.41 &  0.017 & 0.01 & 0.003  &\\
\hline 
\multicolumn{7}{c}{\bf Running parameters} \\
\hline 
$\mu$ & $\mu_\eta$ &  $g_B$ & $\tilde{g}$ &   \\
613 & 395 & 0.0036 & 0.0026 &  \\
\hline 
\multicolumn{7}{c}{\bf Masses} \\
\hline 
$h_1$ & $h_2$ & $h_3$ & $h_4$ & $A^0_1$ & $A^0_2$ &\\
1.4 & 122.4 & 4236 & 4640 &  4264 & 4640 &\\[3mm]
$\tilde{t}_1$ & $\tilde{t}_2$ & $\tilde{q}_{1,2}$ & $\tilde{G}$ & $\tilde{l}$ &  $\tilde{\nu}_R$ &\\
2063 & 3024 & $\sim$4700 & 1750 & $\sim$4500 & $\sim$2000&\\[3mm]
$\tilde{\chi}^0_1$ & $\tilde{\chi}^0_{2,3}$ & $\tilde{\chi}^0_4$ & $\tilde{\chi}^0_5$ & $\tilde{\chi}^0_{6}$ & $\tilde{\chi}^0_7$ & \\
284 & 397 & 531 &  611 & 657 & 680 &\\
\hline 
\multicolumn{7}{c}{\bf Fine-Tuning} \\
\hline 
$\Delta^Z_{m_0}$ & $\Delta^Z_{M_{1/2}}$ & $\Delta^Z_{\mu}$ & $\Delta^{Z'}_{m_0}$ & $\Delta^{Z'}_{\mu_\eta}$ &\\
83 & 172 & 45 & 3347 & 3695 &\\
 \hline 
\hline
\end{tabular}
\caption{An example for a parameter point explaining the $^8\text{Be}$ anomaly within the BLSSM. All dimensionful parameters are given in units of GeV.}
\label{tab:BP}
\end{table}

\section{Conclusion}
\label{sec:conclusion}
In this paper, we have considered the naturalness in supersymmetric models with a light $Z'$ gauge boson. We have shown that the additional fine-tuning due to the new sector can be much smaller than the expected value $O(M_{\rm SUSY}^2/m_{Z'}^2)$. Two mechanisms are used to reduce the fine-tuning. First, it is possible to find relations of the SUSY breaking parameters at the GUT scale for which the a focus point in the MSSM and in the new sector co-exist. We call this Double Focus Point Supersymmetry. Second, we have discussed the importance of radiative corrections to the fine-tuning calculation which can alter the prediction by many orders of magnitude. For both effects one needs a large Yukawa-like ($Y$) coupling to the scalars which are responsible for the gauge symmetry breaking of the new group. We have discussed this explicitly at the example of the $U(1)_{B-L}$ extended MSSM (BLSSM). In particular, the radiative corrections  alter the prediction of the fine-tuning by a factor $Y^4/g^2$, where $g$ is the new gauge coupling. We have confirmed this analytical estimate by a fully numerical calculation. We found that for $m_{Z'}=100$~MeV an additional fine-tuning of $O(100)$ is easily possible for $g \simeq 10^{-4}$. Finally, we have considered the $^8\textrm{Be}$ anomaly in this model. Because of the necessary coupling strength to explain this excess in the BLSSM, a fine-tuning with respect to $Z'$ above $10^3$ seems to be unavoidable as long as perturbativity up to $M_{\rm GUT} \sim 10^{16}$~GeV is demanded. \\
We expect that similar features are present in other SUSY models with extended gauge sectors which involve potentially large Yukawa-like couplings to the new scalars like left-right models \cite{Hirsch:2011hg,Hirsch:2012kv} or separately gauged baryon and lepton number \cite{FileviezPerez:2010gw}.

\section*{Acknowledgements}
We thank Robert Ziegler for interesting discussions, and Chuang Li for technical support. FS is supported by ERC Recognition Award ERC-RA-0008 of the Helmholtz Association.

\begin{appendix}
\section{NBLSSM with small mixing}
\label{app:nblssm}
We now turn to calculate the DFP for $m_{H_u}^2$ and $m_{\eta_2}^2$ in N-BLSSM. After taking the gauge coupling and Yukawa coupling $Y_{\nu}$ to zero, the total one-loop beta functions for $m_{H_u}^2$ and $m_{\eta_2}^2$ are given as
\begin{align}
\beta_{m_{H_u}^2}&=\frac{1}{16\pi^2}\left(6 m_{H_u}^2 y_t^2+6 m_{q}^2 y_t^2+6m_{u}^2 y_t^2
   +6 T_t^2\right)\nonumber\\
\beta_{m_{\eta_2}^2}&=\frac{1}{16\pi^2}\left(2 m_{\eta_2}^2 \lambda^2+ 2m_{\eta_1}^2 \lambda^2+2 m_{s}^2 \lambda^2
   +4 T_{\lambda}^2\right)
\label{eqn:betaone}
\end{align}
where $T_t=y_t A_t$ and $T_{\lambda}=\lambda A_{\lambda}$ are the soft trilinear terms. The calculation of the MSSM section
is completely analog to the BLSSM. The matrix form of the RGEs in the B-L sector are
\begin{align}
\frac{d}{dt}\left[
              \begin{array}{c}
                m_{\eta_2}^2 \\
                m_{\eta_1}^2 \\
                m_{s}^2 \\
                A_{\lambda}^2
              \end{array}
            \right]
            =\frac{\lambda^2}{8\pi^2}
            \left[
                    \begin{array}{cccc}
                      1 & 1 & 1 & 1 \\
                      1 & 1 & 1 & 1 \\
                      1 & 1 & 1 & 1 \\
                      0 & 0 & 0 & 6 \\
                    \end{array}
            \right]
            \left[
              \begin{array}{c}
                m_{\eta_2}^2 \\
                m_{\eta_1}^2 \\
                m_{s}^2 \\
                A_{\lambda}^2 \\
              \end{array}
            \right]~.
\end{align}

This can be solved as
\begin{align}
\left[
              \begin{array}{c}
                m_{\eta_2}^2 \\
                m_{\eta_1}^2 \\
                m_{s}^2 \\
                A_{\lambda}^2 \\
              \end{array}
            \right]
            &=
            \lambda_{6}\left[
              \begin{array}{c}
                1 \\
                1 \\
                1 \\
                3 \\
              \end{array}
            \right]\operatorname{e}^{6J[t]}
           +
            \lambda_{3}\left[
              \begin{array}{c}
                1 \\
                1 \\
                1 \\
                0 \\
              \end{array}
            \right]\operatorname{e}^{3J[t]}\nonumber\\
           &+
           \lambda_{0}\left[
              \begin{array}{c}
                -1 \\
                0 \\
                1 \\
                0 \\
              \end{array}
            \right]
            +
            \lambda_{0}^{\prime}\left[
              \begin{array}{c}
                -1 \\
                1 \\
                0 \\
                0 \\
              \end{array}
            \right]~.
\end{align}

The interesting point is that the exponent $\operatorname{e}^{6I[t]}$ remains approximately $1/3$ even though considering the extended gauge group. Furthermore the exponent $\operatorname{e}^{3J[t]}$ is approximately $4/5$. For the derivation of $\operatorname{e}^{6I[t]}$ and $\operatorname{e}^{3J[t]}$, see appendix~\ref{sec:app}. The subtle point is that the $\operatorname{e}^{6I[t]}$ and $\operatorname{e}^{3J[t]}$ are only calculable when they become Bounlli type. The price we should pay is the smallness of gauge-kinetic mixing couplings i.e. $g_{BY}\sim g_{YB}\sim \mathcal{O}(10^{-2})$.  For now we have
\begin{align}
\left[\begin{array}{c}
                m_{H_u}^2[Q_{\text{GUT}}] \\
                m_{q}^2[Q_{\text{GUT}}] \\
                m_{u}^2[Q_{\text{GUT}}] \\
                A_{t}^2[Q_{\text{GUT}}]
              \end{array}\right]
            &  =\left[\begin{array}{c}
                m_{0}^2 \\
                m_{0}^2+\kappa _0^{\prime}-\frac{2 \kappa
   _{12}}{3} \\
                m_{0}^2-\kappa _0^{\prime}-\frac{4 \kappa
   _{12}}{3} \\
                6\kappa_{12}
              \end{array}\right]\nonumber\\
&\rightarrow\nonumber\\
\left[\begin{array}{c}
                m_{H_u}^2[Q_{\text{SUSY}}] \\
                m_{q}^2[Q_{\text{SUSY}}] \\
                m_{u}^2[Q_{\text{SUSY}}] \\
                A_{t}^2[Q_{\text{SUSY}}]
              \end{array}\right]
              &=\left[\begin{array}{c}
                0 \\
                \frac{m_{0}^2}{3}+\kappa _0^{\prime}-\frac{2 \kappa
   _{12}}{5} \\
                \frac{2m_{0}^2}{3}+\kappa _0^{\prime}-\frac{4 \kappa
   _{12}}{5} \\
                \frac{2}{3}\kappa_{12}
              \end{array}\right]~,
              \end{align}
as well as
\begin{align}
\left[\begin{array}{c}
                m_{\eta_2}^2[Q_{\text{GUT}}] \\
                m_{\eta_1}^2[Q_{\text{GUT}}] \\
                m_{s}^2[Q_{\text{GUT}}] \\
                A_{\lambda}^2[Q_{\text{GUT}}]
              \end{array}\right]
            &  =\left[\begin{array}{c}
                m_{0}^2 \\
                5m_{0}^2+\lambda _0^{\prime}-\frac{4 \lambda
   _6}{5} \\
                9m_{0}^2-\lambda _0^{\prime}-\frac{8 \lambda
   _6}{5} \\
                3\lambda_6
              \end{array}\right]\nonumber\\
 &\rightarrow\nonumber\\
    \left[\begin{array}{c}
                m_{\eta_2}^2[Q_{\text{SUSY}}] \\
                m_{\eta_1}^2[Q_{\text{SUSY}}] \\
                m_{s}^2[Q_{\text{SUSY}}] \\
                A_{\lambda}^2[Q_{\text{SUSY}}]
              \end{array}\right]
             & =\left[\begin{array}{c}
                0 \\
                4m_{0}^2+\lambda _0^{\prime}-\frac{4 \lambda
   _6}{5} \\
                8m_{0}^2-\lambda _0^{\prime}-\frac{8 \lambda
   _6}{5} \\
                \frac{25}{48}\lambda_6
              \end{array}\right]~.
\label{eqn:focus2}
              \end{align}

\section{Appendix}
\label{sec:app}

We show in this appendix the derivation of $\operatorname{e}^{6I[t]}$ and $\operatorname{e}^{3J[t]}$ which are the integral between GUT and SUSY scale for coupling $y_t$ and $\lambda$.
\begin{align}
\operatorname{e}^{6I[t]}&=\exp\left(6\int_{\log Q}^{\log Q_0}\frac{y_t[\rho]^2}{8\pi^2} d\log\rho\right)~,\\
\operatorname{e}^{3J[t]}&=\exp\left(3\int_{\log Q}^{\log Q_0}\frac{\lambda[\rho]^2}{8\pi^2} d\log\rho\right)~.
\end{align}

Since $y_t$ and $\lambda$ are the dominant couplings in the model, their beta functions belong to the Bernoulli type as long as the kinetic mixing couplings become negotiable compared with other gauge couplings. This is subtle in realization of protophobic vector boson model but is suitable for SIDM with light hidden photon mediator.

\begin{align}
\frac{d\alpha_a}{dt}&=2b_a\alpha_a^2~,\nonumber\\
\frac{d\alpha_t}{dt}&=2(s\alpha_t-\sum_a r_a\alpha_a)~,\nonumber\\
\frac{d\alpha_{\lambda}}{dt}&=2(s\alpha_{\lambda}-\sum_a r_a\alpha_a)~,
\label{eqn:Bou}
\end{align}
with the definition $\alpha_t=y_t^2/16\pi^2$,\,$\alpha_a=g_a^2/16\pi^2$ and $\alpha_{\lambda}=\lambda^2/16\pi^2$. Here $s=6$ for $y_t$ and $s=3$ for $\lambda$. $a$ ranges from $1$ to $4$. There is a formal solution for eq.~(\ref{eqn:Bou}),
\begin{align}
\alpha_t[Q]&=\frac{\alpha_t[Q_0]\mathcal{E}_t[Q]}{1-2s\alpha_t[Q_0]\mathcal{F}_t[Q]}~,\nonumber\\
\alpha_{\lambda}[Q]&=\frac{\alpha_{\lambda}[Q_0]\mathcal{E}_{\lambda}[Q]}{1-2s\alpha_{\lambda}[Q_0]\mathcal{F}_{\lambda}[Q]}~,
\end{align}
where
\begin{align}
\mathcal{E}[Q]&=\prod_a\left(1-2b_a\alpha_a[Q_0]\log\left(\frac{Q}{Q_0}\right)\right)
^{\frac{r_a}{b_a}}~,\nonumber\\
\mathcal{F}[Q]&=\int_{\log Q_0}^{\log Q}\mathcal{E}[\rho]d\log \rho~.
\label{eqn:EF}
\end{align}
$r_a$ and $b_a$ can be extracted from the beta function of corresponding Yukawa coupling. Different $\mathcal{E}$ and $\mathcal{F}$ correspond to different $r_a$ and $b_a$. Substituted eq.~\ref{eqn:EF} into the definition of the exponent gives
\begin{align}
\operatorname{e}^{sI[t]}&=\exp\left[2s\int_{\log[Q_0]}^{\log[Q]}
\alpha_t [\rho] d\log[\rho]\right]\nonumber\\
&=\frac{1}{2s\alpha_t[Q_0]\mathcal{F}_t[Q]}\nonumber\\
&=1+\frac{2s\alpha_t[Q]\mathcal{F}_t[Q]}{\mathcal{E}_t[Q]}~,\\
\operatorname{e}^{sJ[t]}&=\exp\left[2s\int_{\log[Q_0]}^{\log[Q]}
\alpha_{\lambda} [\rho] d\log[\rho]\right]\nonumber\\
&=\frac{1}{2s\alpha_{\lambda}[Q_0]\mathcal{F}_{\lambda}[Q]}\nonumber\\
&=1+\frac{2s\alpha_{\lambda}[Q]\mathcal{F}_{\lambda}[Q]}{\mathcal{E}_{\lambda}[Q]}~.
\end{align}

As a consequence, the ratio $\mathcal{F}/\mathcal{E}$ determining the location of focus point. Though $\mathcal{F}$ and $\mathcal{E}$ are varied with the extended gauge group, the ratio remains invariant which is proven numerically. In our case when the strict gauge coupling unification is imposed, we have $\mathcal{E}_t[174]=13.5$ and $\mathcal{F}_t[174]=-135.721$ where the top quark pole mass is chosen as low energy scale. Then the exponent of $\exp(6I[t])$ is approximately $1/3$ which is the same as literature. For $\exp(3J[t])$, we not only need the ratio $\mathcal{F}_{\lambda}/\mathcal{E}_{\lambda}$ but the input value of $\lambda$ at low scale. In order to escape the dangerous landau pole for $\lambda$, it is natural to set $\lambda=0.5$ at low scale. Thus the exponent $\exp(3J[t])$ is approximately $4/5$. The same procedure can be applied to BLSSM with large $\tan\beta^{\prime}$, the exponent $\exp(10K[t])$ is $1/10$.
\end{appendix}

\bibliography{lit}

\end{document}